\documentclass[%
 reprint,
 amsmath,amssymb,
 aps,
]{revtex4-2}
\usepackage{amsmath, wasysym}
\usepackage[toc]{appendix}
\usepackage{graphicx}
\usepackage[caption=false]{subfig}
\usepackage{dcolumn}
\usepackage{comment}
\usepackage{bm}
\usepackage{float}
\usepackage[dvipsnames]{xcolor}
\usepackage{mathcomp}
\usepackage{siunitx}

\newcommand{\rev}[1]{#1}

\usepackage{booktabs}

\usepackage[hidelinks]{hyperref}

\begin{document}
\preprint{APS/123-QED}

\title{Feasibility study of true muonium observation with the existing Belle-II dataset}

\author{Ruben Gargiulo}%
\email{ruben.gargiulo@uniroma1.it}
\affiliation{Università degli Studi La Sapienza, Piazzale Aldo Moro 5, 00185 Roma, Italy \\ INFN Sezione di Roma, Piazzale Aldo Moro 5, 00185 Roma, Italy}%

\author{Elisa Di Meco}
 \affiliation{INFN Laboratori Nazionali di Frascati, Via Enrico Fermi 54, 00044 Frascati, Italy}

\author{Stefano Palmisano}
\affiliation{Università degli Studi La Sapienza, Piazzale Aldo Moro 5, 00185 Roma, Italy \\ Galileo Galilei Institute for Theoretical Physics, Largo Enrico Fermi 2, I-50125 Firenze, Italy}

\newcommand{\spnote}[1]{{\color{Mahogany} #1}} 

\date{\today}

\begin{abstract}
True muonium ($\mu^+\mu^-$) is one of the cleanest bound states, being composed only of leptons, along with true tauonium and positronium. Unlike the latter, true muonium and true tauonium have not been observed so far.
This article shows that the spin-0 state of true muonium (para-TM), decaying into two photons, can be observed at a discovery level of significance in the dataset already collected by the Belle-II experiment at the $\Upsilon (4S)$ peak, with certain assumptions on systematic uncertainties.
Para-TM is
produced
via photon-photon fusion, and its observation is based on the detection of the photon pair resulting from its decay, on top of the continuum background due predominantly to light-by-light scattering. 
Trigger, acceptance and isolation cuts, along with calorimeter resolution and reconstruction efficiency, are taken into account during the Monte Carlo simulation of both signal and background. In order to separate signal and background, a machine learning method, based on extremely randomized trees, is trained on simulated events. Finally, the expected statistical significance of TM observation is evaluated, taking into account systematic uncertainties in a parametric fashion.

\end{abstract}

\maketitle


\section{\label{sec:intro} Introduction}

\begin{figure}[htpb]
    \includegraphics[width=0.7\linewidth]{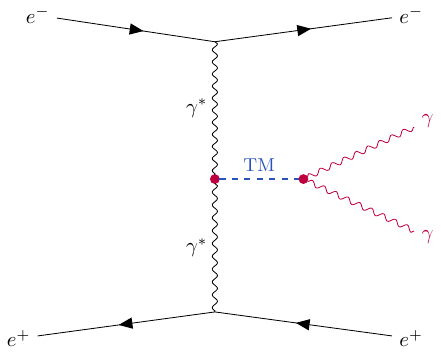}
    \caption{Feynman diagram for $e^+ e^- \to e^+ e^- \gamma \gamma \to e^+ e^- TM$, including the TM decay into two photons.}
    \label{fig:feynTM}
\end{figure}

Quantum electrodynamics (QED) predicts the existence of several bound states, in addition to standard atoms, such as purely leptonic systems.
The lightest one, positronium ($e^+e^-$), has been discovered decades ago and extensively studied \cite{positronio}. In contrast, true muonium ($\mu^+\mu^-$) and true tauonium ($\tau^+\tau^-$) have never been observed.
Regarding true muonium (TM), several methods have been pointed out for its observation, but none has been accomplished so far.
Some of these proposals require building or modifying $e^+ e^-$ colliders \cite{dimus} \cite{gargiulo2024true} \cite{russi}, constructing detectors in existing beam facilities \cite{h4} \cite{Holvik:1986ty}  \cite{sakimoto}, or exploiting future datasets from running colliders \cite{Brodsky:2009gx} \cite{lhcb} \cite{Ginzburg:1998df}.
In this paper, a novel method for observing TM is described, requiring the analysis of existing data already collected by the Belle-II experiment \cite{belle2} \rev{in $e^+ e^-$ collisions} at the $\Upsilon(4S)$ peak, with $\sqrt{s} = 10.58$ GeV. True muonium (TM) should be observed in its spin-0 fundamental state (para-TM), decaying into two photons. Its production is accomplished via photon-photon fusion ($e^+ e^- \to e^+ e^- \gamma \gamma \to e^+ e^- TM$), as sketched in Figure \ref{fig:feynTM}.

The electron and the positron in the final state scatter at very low angles and are not expected to be detected, therefore the experimental signature consists of two isolated photons with a combined invariant mass around TM mass, i.e. $m_{\gamma \gamma} = m_{\mathrm{TM}} \sim 2m_{\mu} \sim 0.211$ GeV. A veto on any high-energy lepton in detector acceptance must be applied: this is possible thanks to the small $O(10^{-3})$ events/bunch crossing physics pile-up in the Belle-II environment, with $O(100)$ MHz bunch crossing frequency, $O(300)$ nb total cross section (dominated by Bhabha scattering) in acceptance and $8\times 10^{35}$cm$^{\rev{-2}}$s$^{-1}$ maximum instantaneous luminosity. 

\begin{figure}[H]
\centering
\includegraphics[width=.9\linewidth]{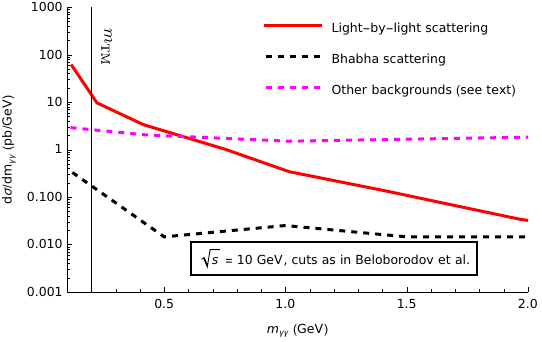}
\caption{Differential cross-sections $\frac{d \sigma}{d m_{\gamma \gamma}}$ 
for the production of a photon pair with photon polar angles between 30$^\circ$ and 150$^\circ$ and a $m_{\gamma \gamma} $ invariant mass. 
At $m_{\gamma\gamma} = m_{\text{TM}}$ relevant to this work, light-by-light scattering (red curve, from \cite{russi2}) constitutes the largest background, while other backgrounds such as $e^+e^-\to e^+e^-\gamma\gamma\gamma$, $e^+e^-\to \gamma\gamma\gamma\gamma$, $e^+e^-\to e^+e^-\gamma\gamma\gamma\gamma$ (magenta curve, from \cite{russi2}), and doubly-radiative Bhabha scattering (black curve) are subdominant. The latter was simulated using Madgraph5 \cite{madgraph} with cuts compatible with those of \cite{russi2}.
}
    \label{fig:xsbkg}
\end{figure}

At the Belle-II center-of-mass energy, the dominating background, for $\gamma \gamma$ final states with $m_{\gamma \gamma} \sim 0.211$ GeV, is light-by-light scattering. 
Other minor backgrounds are represented by $e^+e^- \to 4\gamma$ (roughly one order of magnitude smaller), i.e. $e^+ e^- \to \gamma \gamma$ with the emission of two undetected photons, and by $e^+ e^- \to e^+ e^- 3\gamma$ (two order of magnitudes smaller). 
The background $e^+e^-\gamma \gamma$, with both leptons undetected could also arise from double-radiative Bhabha scattering, which is not fully included in \cite{russi2}, therefore it was simulated with Madgraph5 \cite{madgraph} with generation cuts 
compatible with those of \cite{russi2}. Results are shown in Fig. \ref{fig:xsbkg}. 


Backgrounds due to the beam are not taken into account in this paper.

As the discovery potential is unlikely to be undermined by neglecting minor backgrounds with the same signatures as the dominant one, for simplicity in this paper only the main background is taken into account, i.e. light-by-light scattering. However, in the case of a real experimental analysis, when data and simulated samples are required to match, it is strongly suggested to include all three processes presented above.

In the following sections, the cross-sections for signal and light-by-light scattering are evaluated analytically and compared to the results of SuperChic \cite{superchic}, a Monte Carlo event generator. Samples from SuperChic for both signal and backgrounds are then processed including experimental cuts, resolutions and efficiencies and then fed into a machine learning model based on extremely randomized trees (ExtraTrees) \cite{extratrees}, to separate them. In the final section, the discovery significance is evaluated for the existing Belle-II dataset, considering also the effect of systematic uncertainties.

\section{Cross-sections evaluations and event generation}
Both TM production and light-by-light scattering arise from 
the interactions of intermediate virtual photons.
Hence, they can be factorized as the production of two photons  from the colliding $e^+e^-$ pair, followed by the hard interactions $\gamma \gamma \to TM$, or $\gamma \gamma \to \gamma \gamma$.

For a given hard process $\gamma \gamma \to X$ with a final state $X$ (TM or $\gamma \gamma$ in the cases of interest), the cross-section is evaluated using the equivalent photon approximation (EPA) \cite{a29}, 
which consists in treating the intermediate photons as real -- namely on shell and transversly polarized -- and considering the cross-section for 
the real process $\gamma \gamma \to X$
 along with the corresponding photon fluxes of the colliding $e^+ e^-$ pair \cite{tt}.
The differential cross-section $\frac{d \sigma(e^+e^- \to e^+ e^- \gamma \gamma \to e^+e^- X)}{d m_{\gamma \gamma}}$ with respect to the final-state invariant mass $m_{\gamma \gamma}$ is 
computed as:
\begin{equation}
    \frac{d \sigma}{d m_{\gamma \gamma}} = \sigma(\gamma \gamma \to X)\big|_{\sqrt{s}_{\gamma \gamma} = m_{\gamma \gamma}} \frac{d \mathcal{L}_{\gamma \gamma}}{d W_{\gamma \gamma}}\bigg|_{W_{\gamma \gamma} = m_{\gamma \gamma}}
    \label{eq:dxsdm}
\end{equation}
where $\frac{d \mathcal{L}_{\gamma \gamma}}{d W_{\gamma \gamma}}$ is the value of the effective two-photon luminosity function at the resonance mass, computed from the two incoming photon EPA fluxes \cite{russi2}.

\subsection{Equivalent two-photon luminosity function}

The photon flux for each lepton beam is estimated
with the Weizsäcker-Williams approximation \cite{a30}.
Hence the spectrum of colliding equivalent photons is given by the formula $dn_{\gamma} = f(x) dx$:
    \begin{equation}
    \begin{split}
    f(x) = \frac{\alpha}{2\pi} \bigg[ & 
\frac{1 + (1 - x)^2}{x}{\ln{\frac{Q_{\text{max}}^2}{Q_{\text{min}}^2(x)}}} 
+ \\
& 2m_e^2 x \left( \frac{1}{Q_{\text{max}}^2} - \frac{1}{Q_{\text{min}}^2(x)} \right) 
\bigg]
\end{split}
\end{equation}
where $Q^2_{\text{min}}(x) = m^2_e x^2 / (1 - x)$  and $x$ is the fraction of the beam particle momentum taken by the photon (
cf. Eq. (3) of \cite{a31}).
Note that during computations, when $Q^2_{\text{min}}(x) > Q^2_{\text{max}}$, which happens for some values of $x$, $Q^2_{\text{min}}(x)$ is imposed to be equal to $Q^2_{\text{max}}$.
As in the EPA the two photons with relative momenta $x_1 , \, x_2$ are approximated to be collinear to the beam particles, the two-photon invariant mass is evaluated as $W_{\gamma \gamma} = \sqrt{s} x_1 x_2$. If $x_1 = x$, then $x_2 = z^2/x$, where $z = W_{\gamma \gamma} / \sqrt{s}$, therefore the effective two-photon luminosity differential is evaluated as following \cite{russi2}:
\begin{align}
    d \mathcal{L}_{\gamma \gamma} = \int dn_1 dn_2 = \int f(x_1) f(x_2)\, dx_1\, dx_2 = \\ = 2z \, dz \int f(x) f \left ( z^2/x \right ) dx / x
\end{align}
and:
\begin{equation}
    \frac{d \mathcal{L}_{\gamma \gamma}}{d W_{\gamma \gamma}} = \frac{2z}{\sqrt{s}} \int f(x) f \left (\frac{z^2}{x} \right) \frac{dx}{x}\,.
\end{equation}
 The maximum virtuality is set to
$Q^2_{\mathrm{max}} = m_{\mathrm{TM}}^2 / 10$, in order to ensure that in the analyzed kinematical regime the EPA is valid. The same cut is also employed in the simulation of events. In this way, cross-sections are actually bounded from below, but one can estimate the accuracy of the EPA in this case as 
$\frac{Q^2_{\text{max}}}{m_{\text{TM}}^2 \log \left(\frac{Q^2_{\text{max}}}{Q^2_{\text{min}}}\right)}\sim \rev{0.1 \%}$ 
for an average $x^2 \sim m_{\mathrm{TM}}/\sqrt{s}$ \cite{a29}. 

The resulting value of $\frac{d \mathcal{L}_{\gamma \gamma}}{d W_{\gamma \gamma}}$ at the TM mass is 0.0816 GeV$^{-1}$.

\subsection{Signal}
Para-TM decays in two photons with a very small width $\Gamma_{\mathrm{TM}}^{\gamma \gamma} = \alpha^5 m_\mu / 2 \sim 1.0931 \times 10^{-12}$ GeV, therefore to evaluate the production cross-section $\sigma_{\mathrm{TM}} = \sigma(e^+e^- \to e^+ e^- \gamma \gamma \to e^+e^- TM)$
the narrow-width approximation can be employed, and
only the value of $\frac{d \mathcal{L}_{\gamma \gamma}}{d W_{\gamma \gamma}}$ at $W_{\gamma \gamma} = m_{\mathrm{TM}}$ is required
\cite{tt}:
\begin{equation}
    \sigma_{\mathrm{TM}} = 4 \pi^2 \frac{\Gamma_{\mathrm{TM}}^{\gamma \gamma}}{m_{\mathrm{TM}}^2} \frac{d \mathcal{L}_{\gamma \gamma}}{d W_{\gamma \gamma}} \bigg|_{W_{\gamma \gamma} = m_{\mathrm{TM}}} \sim 30.6 \text{ fb}
    \label{signal}
\end{equation}

The generation of Monte Carlo events for the production and subsequent decay of para-TM into two photons was performed by using the SuperChic code\cite{superchic}, which 
\rev{allows} simulating photon-photon fusion events from lepton beams with the EPA, when photons fuse into resonances. In particular, it features the possibility to generate events with axion-like particles (ALPs) as resonances \cite{alp-lhc}. ALPs have the same quantum numbers $J^{PC} = 0^{-+}$ as TM and also couple to photon pairs both in production and in decay, therefore the angular distribution of photon pairs generated by ALPs decay is the same as the TM, i.e. the one expected for pseudoscalar particles. The photon quadrimomenta distributions are also the same, because with fixed decay angles, photon energies depend only on the resonance quadrimomentum, and both the TM and the ALP quadrimomentum depend only on the EPA photon spectra.
The cross-section for production of ALPs is given by the same general formula (see Eq. 2 in \cite{tt}) as Eq. \eqref{signal}, then, to get correct cross-sections values in SuperChic, the ALP mass is set to the TM mass, and the coupling parameter $g_a$ is set to a value such that the ALP width 
\begin{equation}
    \Gamma_a = \frac{g_a^2 m_a^3}{64 \pi}.
\end{equation}
matches the TM width, i.e. $g_a^{TM} \sim 1.5 \times 10^{-4} $ GeV$^{-1}$.
This procedure ensures the correct evaluation of signal cross-section and outgoing photons quadrimomenta distributions.

The actual confirmation of the validity of this 
approach comes from the cross-section $\sigma_{\mathrm{TM}} = 30.7 \pm 0.2$ fb, obtained with SuperChic after generating $10^3$ ALP events with the correct coupling and mass and applying the cut $Q^2_{\mathrm{max}} = m_{\mathrm{TM}}^2 / 10$ (ensuring the EPA validity). Note that the value from the SuperChic ALP generation is compatible within uncertainties with the value evaluated analytically. 
Note also that 
this approach
has been further validated, by re-evaluating with SuperChic the true tauonium production cross-section calculated in \cite{tt}. The spin-0 true tauonium (para-TT) production cross-section was evaluated in the same kinematical region ($|\eta| < 5$) 
in
Table II 
of
\cite{tt} and obtaining the same value of $0.015$ fb, after setting $Q^2_{\mathrm{max}}$ to the same value of $1$ GeV$^2$, which is close to$ \frac{m_{\text{TT}}^2}{10}$, thus ensuring the EPA validity.

\subsection{Background}
\begin{figure}[htpb]
    \centering
    \includegraphics[width=0.8\linewidth]{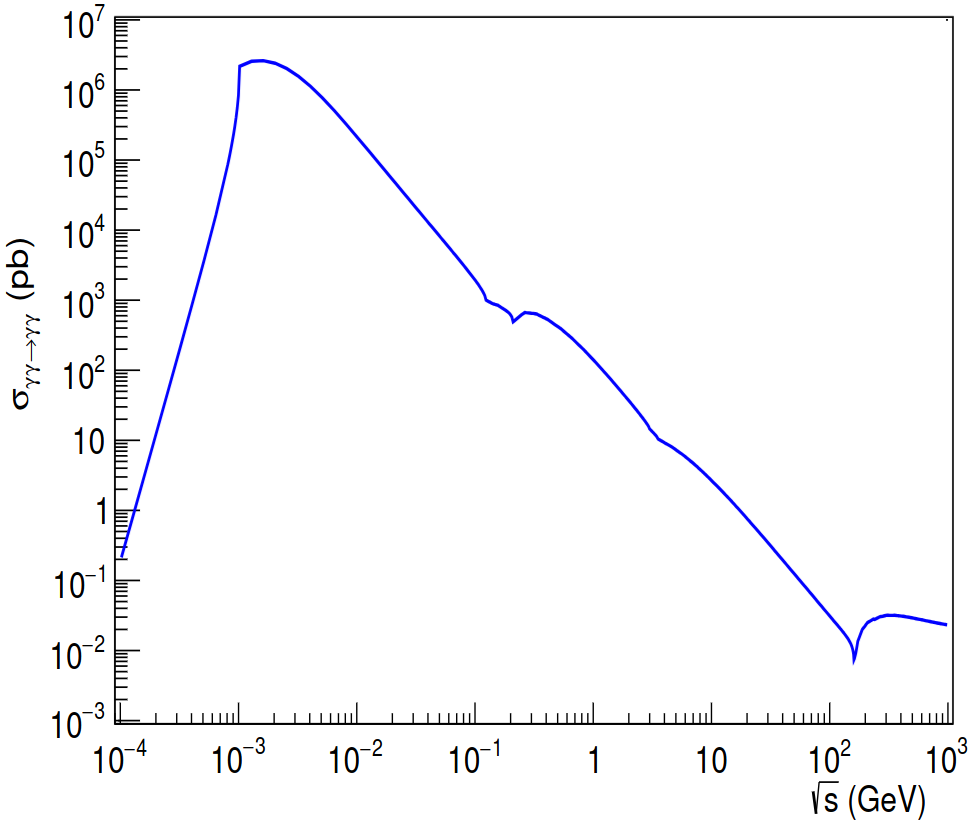}
    \caption{Cross-section of the $\gamma \gamma \to \gamma \gamma$ (from \cite{russi2}) evaluated with the SANC program.}
    \label{fig:gggg}
\end{figure}

As discussed in Sec. \ref{sec:intro}
, the dominant background, and the only one considered in this paper, is the light-by-light scattering. 
The differential cross-section for light-by-light scattering at the TM mass is given by Eq. \eqref{eq:dxsdm}:
\begin{equation}
    \frac{d \sigma}{d m_{\gamma \gamma}} = \sigma(\gamma \gamma \to \gamma \gamma) \frac{d \mathcal{L}_{\gamma \gamma}}{d W_{\gamma \gamma}}\bigg|_{W_{\gamma \gamma} = m_{\gamma \gamma}}
\end{equation}
The cross-section $\sigma(\gamma \gamma \to \gamma \gamma)$ is equal to about 502 pb at the TM mass, from calculations made with the SANC program \cite{sanc}, as shown in Figure \ref{fig:gggg}, translating in a differential cross-section at the TM mass of about 41.85 pb/GeV.
In order to study the TM discovery potential, background events were simulated with SuperChic.
A validation of SuperChic performances for light-by-light scattering generation was performed by comparing the differential cross-section calculated at the TM mass evaluated analytically to the one evaluated by SuperChic. When applying cuts in the two-photons invariant mass in a 200 keV window around the TM mass, a value of 43.0 pb/GeV was obtained, in agreement within 3\% with the number evaluated analytically.
As for the signal, in this paper the background cross-section values used in future calculations will be the one provided by SuperChic.
A further validation was performed by evaluating light-by-light scattering cross-section in a 200 MeV mass window around para-TT in the same kinematical region ($|\eta| < $ 5)
in
Table II 
of
\cite{tt} and obtaining the same value of 1.7 fb, after setting $Q^2_{\mathrm{max}}$ to the same value of $1$ GeV$^2$.

\section{Parametric detector simulation}
The effects of experimental resolution reconstruction and trigger on signal and background events were taken into account, as explained in the following.
The collider asymmetry was also included, by applying a $\beta = 0.28$ boost in the collider axis \cite{belle2-nim}, in order to obtain simulated events in the lab frame.
The Belle-II electromagnetic calorimeter is divided in three parts (barrel, forward endcap, and backward endcap) with different energy resolutions and reconstruction efficiencies
\cite{calognn}.
The effect of the energy resolution is considered, by smearing generated events using the formula $\sigma_E/E = a/E\text{[GeV]} \oplus b/\sqrt{E\text{[GeV]}} \oplus c$. On the contrary, the reconstruction efficiency is approximated to be flat in energy and applied as an event weight, as shown in Table \ref{tab:calo} using data from \cite{calognn}.
\begin{table}[htpb]
\begin{tabular}{lcccccc}
Region & $\theta_{\text{min}}$& $\theta_{\text{max}}$ & a  & b      & c      & $\epsilon$ \\
\hline
Barrel & 32.2  & 128.7 & 0.0125 & 0.0239 & 0.0075 & 0.8  \\
Forward endcap & 12.4  & 31.4  & 0.0061 & 0.0223 & 0.012  & 0.7  \\
Backward endcap & 130.7 & 155.1 & 0.0218 & 0.0251 & 0.0228 & 0.65
\end{tabular}
\caption{Angular regions ($\theta_{\text{lab}}$), parameters for the energy resolution $\sigma_E/E = a/E\text{[GeV]} \oplus b/\sqrt{E\text{[GeV]}} \oplus c$, and average efficiency for the different regions of Belle-II electromagnetic calorimeter \cite{calognn}, as used in this paper to parametrize the detector response.}\label{tab:calo}
\end{table}

An angular resolution of 13 mrad in $\theta$ and $\phi$ \cite{belle2} is also taken into account by smearing generated events, along with isolation cuts ($\Delta \theta > 48$ mrad and $\Delta \phi > 48$) 
between the clusters of the two photons \cite{franceschini, belle2}, and trigger cuts for the two-photon physics path ($\theta_{\text{lab}} \geq  17^\circ, \, p_T \geq 0.1$ GeV) \cite{trigger}.

\section{Background Suppression}

\begin{figure}[htpb]
    \centering
    \includegraphics[width=\linewidth]{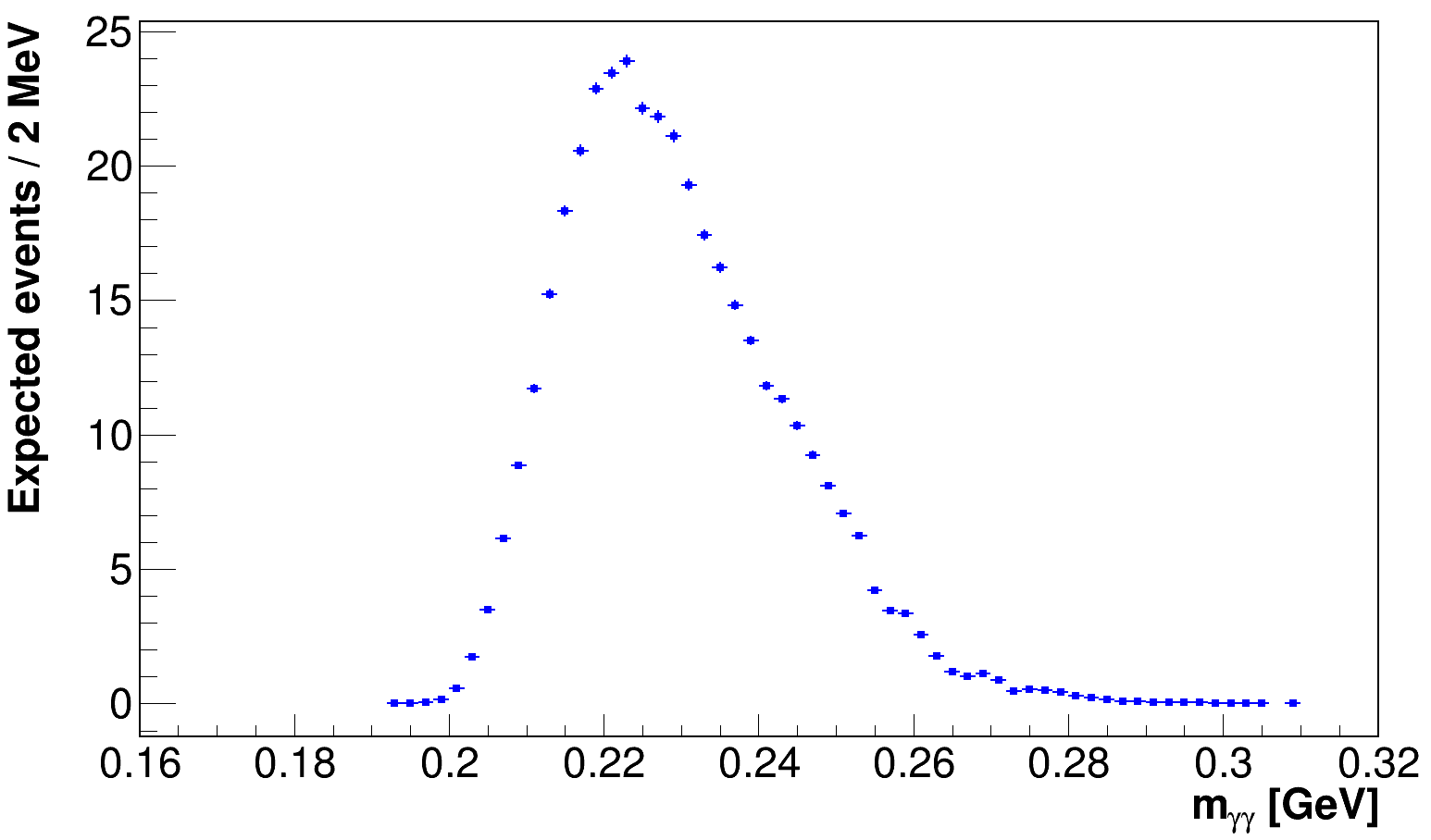}
    \vspace{0.5cm}
    \includegraphics[width=\linewidth]{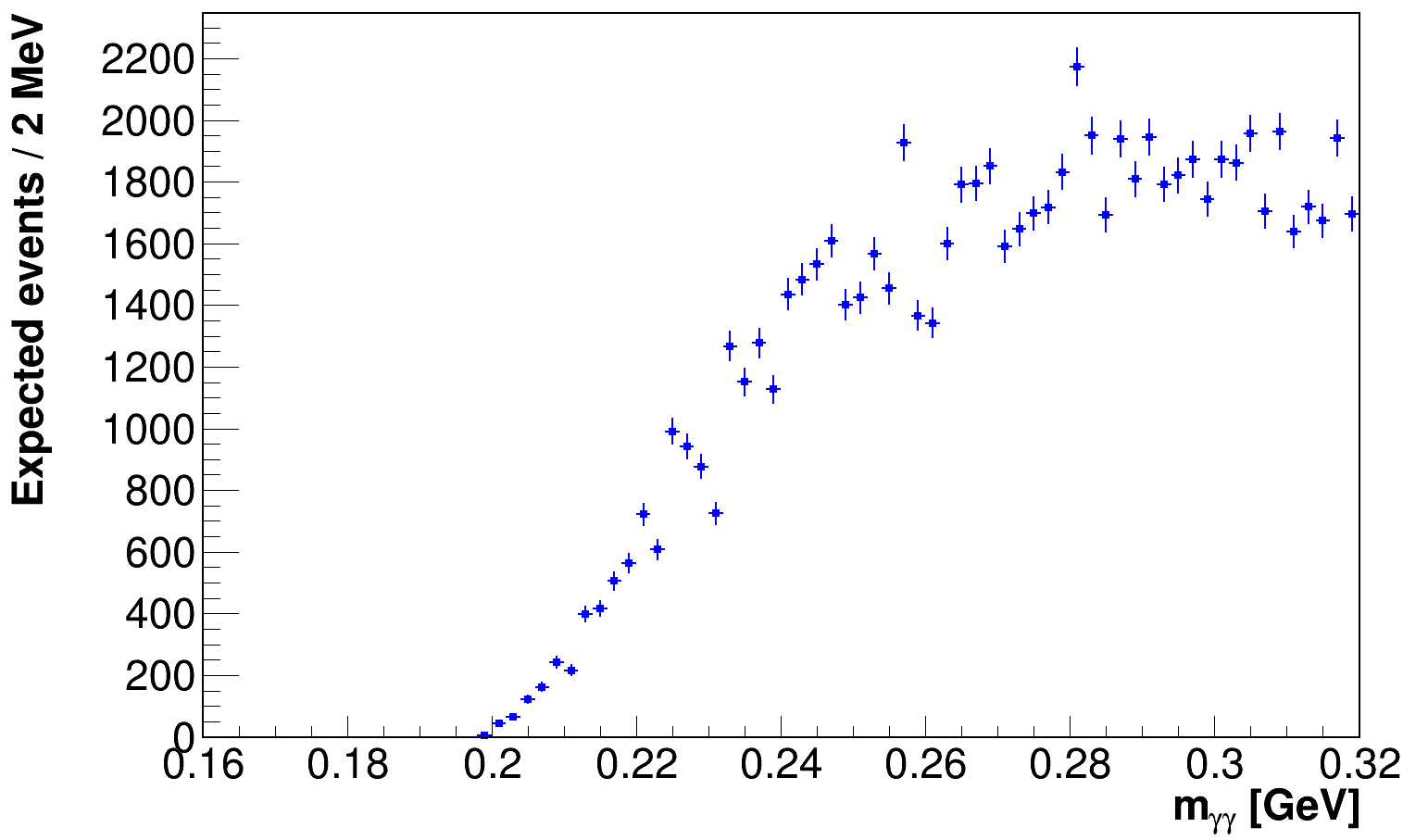}
    \caption{Signal and light-by-light scattering spectra in $m_{\gamma \gamma}$ of simulated samples, with detector and trigger effects taken into account, corresponding to the 363 fb$^{-1}$ luminosity already collected by the Belle-II experiment at the $\Upsilon(4S)$ peak.}
    \label{fig:spectra}
\end{figure}

The spectra in $m_{\gamma \gamma}$ of signal and background simulated samples, including all said detector and trigger effects, corresponding to the 363 fb$^{-1}$ luminosity already collected by the Belle-II experiment \rev{during Run 1 (up to end of 2022)} at the $\Upsilon(4S)$ peak, are shown in Figure \ref{fig:spectra}.\\

The threshold-like shape in both signal and background is due to the combined effect of the cuts on $Q^2_{\mathrm{max}}$ and photons $p_T$, motivated by, respectively, necessity to ensure the EPA validity, and trigger.

If the signal spectrum is integrated in the range [0.195-0.24] GeV, containing about the 80\% of the distribution, a total of about $S=311$ events is collected, while in the same range about $B\sim13100$ background events are integrated. The signal region edges were found as a result of an optimization. The statistical significance without further cuts is $S/\sqrt{B} \sim 2.7\,\sigma$, therefore to reach 5 $\sigma$ a finer analysis is needed.
Actually, in the presence of systematic error on the background yield, the significance $Z$ is lower, and must be evaluated as in the general case:
\begin{equation}
    Z = \frac{S}{\sigma_{B}} = \frac{S}{\sqrt{B + \left ( \frac{\sigma^{\text{syst}}_B}{B} \right)^2 B^2 }}
\end{equation}
where $\sigma^{\text{syst}}_B/B$ is the relative systematic uncertainty on background yield in the signal region.
For example, with systematics uncertainties of 0.5\%, 1\%, 2\% and 5\%, the significance is, respectively, 2.3$\sigma$, 1.8$\sigma$, 1.1$\sigma$, 0.5$\sigma$.

A realistic estimate of $\sigma^{\text{syst}}_B/B$ is very difficult with the information available to the authors
, as it depends heavily on detector effects, but some observations can be made. The most worrying part of the uncertainty likely comes from the knowledge of the ratio between the background yield in the signal region and in the right sideband region. It mainly depends on how finely efficiencies and cross-sections are known, with $m_{\gamma \gamma}$ varying at the scale of tens of MeV. 
It is outside the scope of this paper to comment further, therefore it will be assumed that the relative systematic uncertainty on background yield in the signal region $\sigma^{\text{syst}}_B/B$ is between 0.5\% and 5\% and the final significance will be evaluated for different $\sigma^{\text{syst}}_B/B$ values.
In order to achieve discovery level significances, some kind of background suppression must be implemented. In this work, the choice fell on a machine learning model able to make a multi-variated signal/background discrimination.

\subsection{Extra Trees classifier}
The machine learning model of choice is an Extremely random forest of binary trees (ExtraTrees) \cite{extratrees}.

In standard random forests, each tree in the ensemble is constructed using a sample drawn with replacement from the training set \cite{sklearn}.
Moreover, when splitting a node during tree construction, the optimal split is determined by exhaustively searching the feature values across either all input features or a random subset of fixed size.

In contrast, ExtraTrees forests use a random subset of candidate features, but rather than searching for the most discriminative thresholds, thresholds are randomly selected for each candidate feature, and the best of these randomly generated thresholds is chosen as the splitting rule. This approach often helps to reduce overfitting.

Machine learning models used for classification of high-energy particle physics events often tend to learn the shape of the signal in the most easy variable, in this case $m_{\gamma\gamma}$, and to reproduce this shape for background events classified as signal. This issue, named mass sculpting, would reduce the classification power of the model and potentially spoil the measurement.

\begin{figure}[H]
    \centering
    \includegraphics[width=0.8\linewidth]{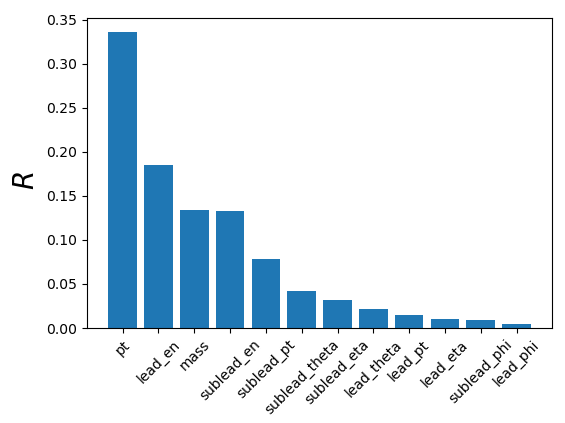}
    \caption{\rev{Pearson correlation coefficient ($R$) between the classifier score and each feature. The \textit{lead} and \textit{sublead} prefixes refer to photons with higher and lower $p_T$, respectively, while the \textit{en, theta, eta, phi} suffixes, refer to energy, polar angle, pseudorapidity and azimuthal angle, respectively. }}
    \label{fig:r2}
\end{figure}

In order to prevent mass sculpting and to force the model to learn background suppression strategies not related to the mass, the signal sample used to train the model was generated by putting together $10^3$ signal samples with ALP masses evenly spaced from 50 MeV to 380 MeV and couplings tuned to have cross-sections equal to the TM one. The resulting spectrum in mass is however increasing because of detector acceptance. The flattened signal sample and half of the generated background sample, with all said detector and trigger effects taken into account, were fed into an ExtraTree classifier with 100 trees and with a fixed size of randomly selected features of 4.  For each event the classifier outputs a number from 0 to 1, discretized in 100 steps (as the number of trees), representing the probability that the event is a signal event.
\rev{The input features include energy, transverse momentum, pseudo-rapidity, azimuthal angle and polar angle for both photons, together with combined mass and transverse momentum. The relative importance of each feature is estimated as shown in Figure \ref{fig:r2}, using as figure of merit the absolute Pearson correlation coefficient between the classifier score and each feature. The feature showing the highest correlation with the classifier score is the diphoton $p_T$.}

\begin{figure}[htpb]
    \centering
    \includegraphics[width=1\linewidth]{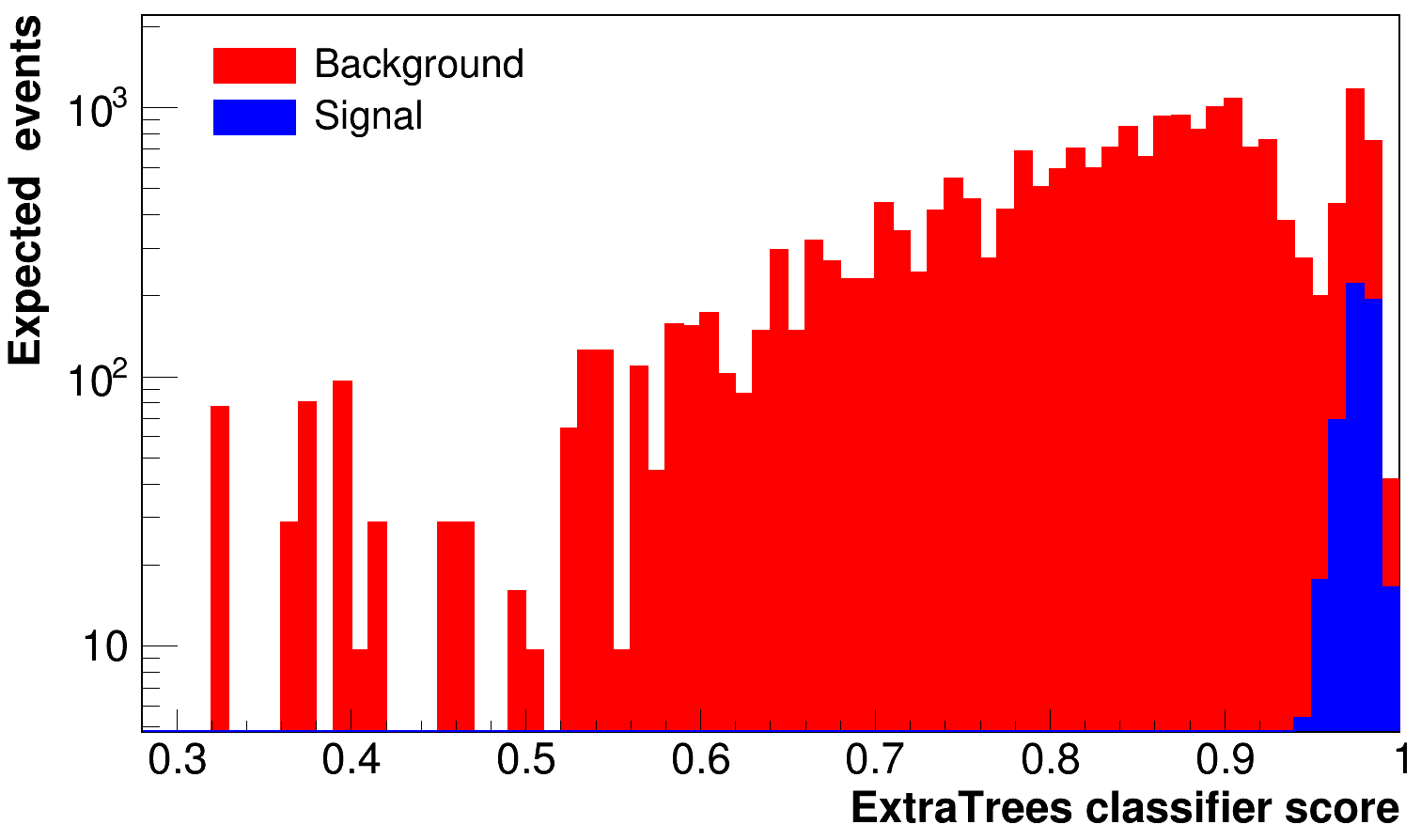}
        \caption{Distribution of ExtraTrees score for signal and light-by-light scattering samples in the signal mass region, with statistics corresponding to the 363 fb$^{-1}$ luminosity already collected by the Belle-II experiment at the $\Upsilon(4S)$ peak.}
    \label{fig:scores}
\end{figure}

After training, the model was evaluated using the original signal sample, i.e. the one generated at the TM mass, which was never fed into the model for training, and the other half of the generated background sample, which also was not used for the training, and the resulting score distributions are shown in Figure \ref{fig:scores}.\\
\ \\

\section{Discovery potential}
\begin{figure}[htpb]
    \centering
    \includegraphics[width=\linewidth]{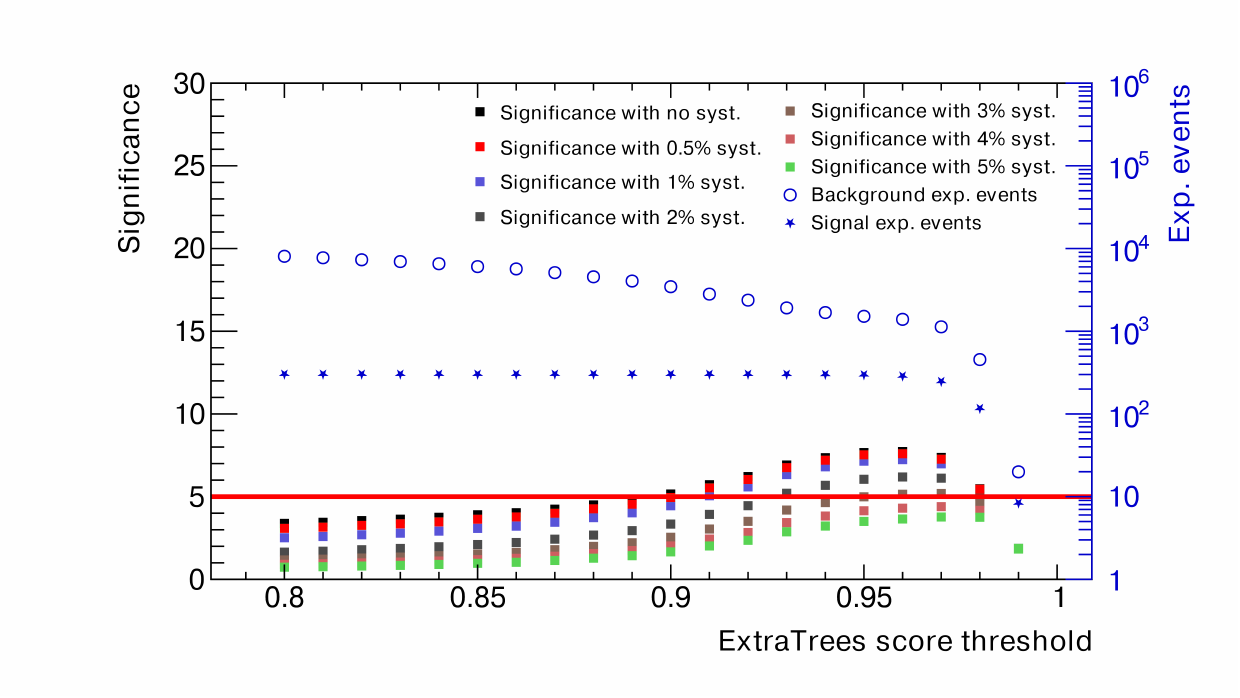}
    \caption{Significance vs. threshold on the ExtraTrees classifier score ($c_t$) for different values of systematics on the background yield in the signal region [0.195, 0.24] GeV, along with the expected number of events for signal and background depending on the threshold.}
    \label{fig:sign}
\end{figure}

\begin{figure}[htpb]
    \includegraphics[width=0.9\linewidth]{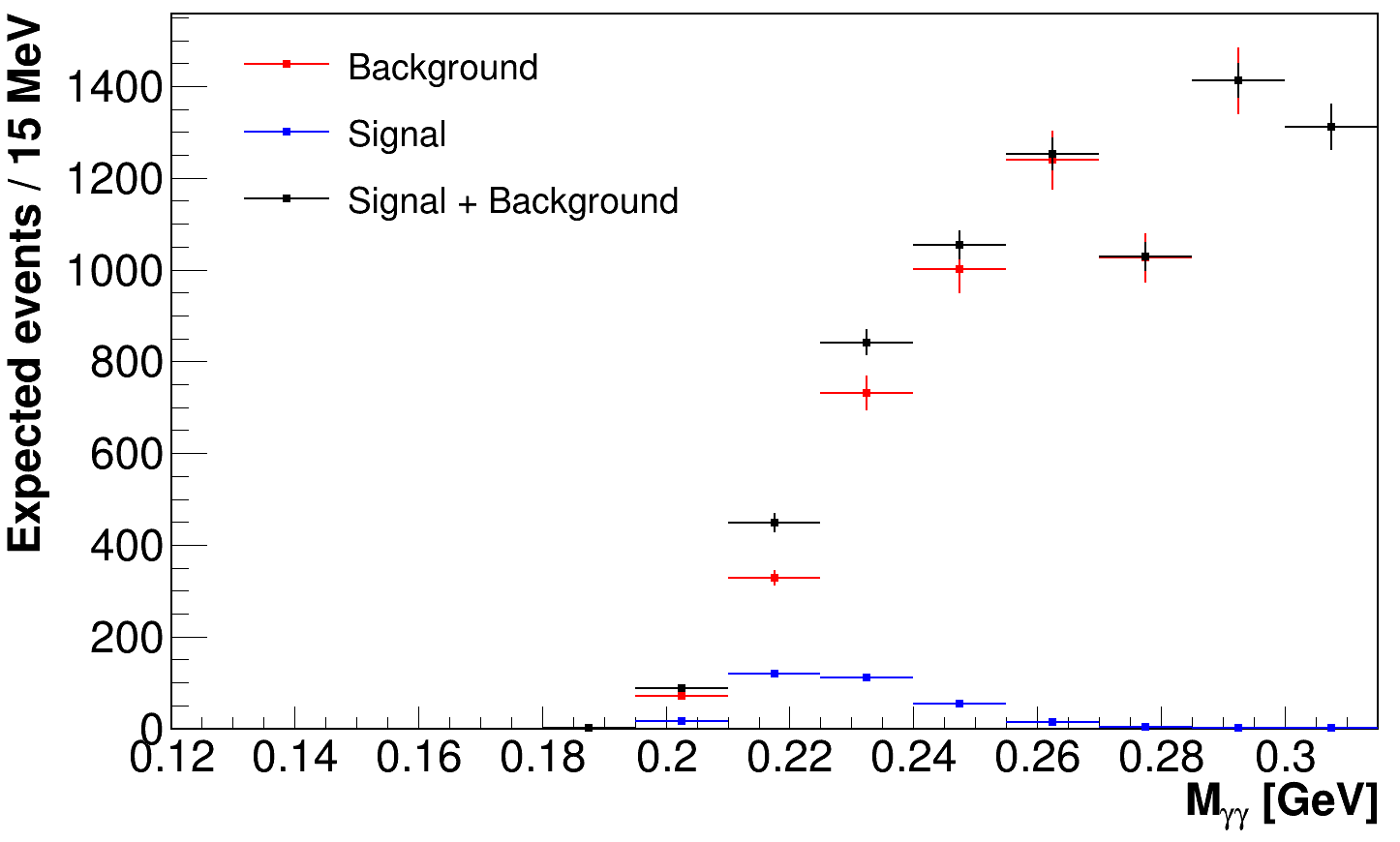}
    \caption{Distributions of $m_{\gamma \gamma}$ for signal, background and their sum for a threshold value of 0.97 and a systematic uncertainty of 3\% on background yield in the signal region. The errors on Signal+Background histogram bins are statistical, while the ones on the background are systematic, and adjusted in a flat manner such that the systematic uncertainty on background yield in the signal region is 3\%.}
    \label{fig:shape}
\end{figure}

The score $c$ of the ExtraTrees classifier was used to discriminate signal over background with a simple threshold $c_t$, i.e. events with $c > c_t$ are classified as signal, while events with $c \leq c_t$ are taken as background.
The $c_t$ value was optimized using the value of the resulting significance with a 3\% systematics on the background yield in the signal region, as shown in Figure \ref{fig:sign}. The significance was also evaluated with systematics of 0\% (no systematics), 0.5\%, 1\%, 2\%, 3\%, 4\%, and 5\%. Results show that the main purpose of the cut on $c_t$ is to suppress the background with a very small score, and the significance is therefore rather flat.
Discovery significances exceeding 5$\sigma$ are achieved for systematics less than or equal to 3\%.
An example of $m_{\gamma \gamma}$ for signal, background and their sum is shown in Figure \ref{fig:shape} for a threshold value of 0.97 and a systematic uncertainty of 3\% on background yield in the signal region. 
The threshold-like shape is due to the combined effect of the cuts on $Q^2_{\mathrm{max}}$ and photons $p_T$, as already discussed above. When analyzing real collider data, the $Q^2_{\mathrm{max}}$ cut could be removed, thus recovering a flatter shape for the background.

It is possible that a better background discrimination strategy could achieve higher significance for all systematics values and discovery significance also for 4\% or 5\% systematics, by means of other machine learning models or analysis cuts.

\subsection{Associated production}
It could also be possible to see hints of TM production in the Belle-II existing dataset by using the production of TM in association with a photon, i.e. the process $e^+e^- \to TM \gamma \to 3 \gamma$, not considered in this paper. 

Expected limits for $e^+e^+ \to a \gamma$, where $a$ is an ALP, are available in literature at different values of integrated luminosity \cite{alpbelle}, and point to 95\% C.L. limits $g_a < 3 \times 10^{-4}$ GeV$^{-1}$ for $m_a = m_{\mathrm{TM}}$ with 20 fb$^{-1}$. Given that the limit scales with the inverse quartic root of luminosity (due to the quadratic scaling of the $e^+e^+ \to a \gamma$ cross section with respect to $g_a$), the expected limit with the existing dataset sets around $1.5 \times 10^{-4}$ GeV$^{-1}$, which is equal to the effective $g_a$ value for TM. Therefore a 2$\sigma$ observation of TM with $e^+e^- \to TM \gamma$ should be possible. Moreover, a combined analysis with photon-photon fusion and this method could improve the significance.\\
\ \\
\subsection{Photon-photon fusion with visible \texorpdfstring{$e^+e^-$}{e+e-} final states}
Also for the case of photon-photon fusion with visible $e^+e^-$ final states, expected limits for ALP production are available in literature \cite{franceschini}. By using again the inverse quartic root scaling of the expected limit with luminosity, a 95\% C.L. limit $g_a < 10^{-5}$ GeV$^{-1}$ for $m_a = m_{\mathrm{TM}}$ with 363 fb$^{-1}$ (the current integrated value) is derived, which is more than 15 times smaller than $g_a^{TM}$. For this reason, this channel could be sensitive to TM at a discovery level of significance.

\section{Conclusions}
This work  is focused on
the potential of observing at discovery level of significance the spin-0 state of True Muonium (para-TM), a bound state of a muon and an anti-muon not yet observed, using existing data taken by the Belle-II experiment \rev{in $e^+ e^-$ collisions} at the $\Upsilon(4S)$ peak. 

The production of para-TM through photon-photon fusion (with undetected final-state leptons) and its subsequent decay into two photons was investigated.

The study involves simulating Monte Carlo events using the program SuperChic, to differentiate between the signal (para-TM decay) and the background, primarily light-by-light scattering. The simulation incorporates experimental effects such as trigger, acceptance, resolutions, and detector efficiency.

In order to enhance the separation of signal and background, a machine learning model based on Extremely Randomized Trees (ExtraTrees) was employed.

It was found that, with systematic uncertainties on background yield in a certain signal region ([0.19, 0.245] GeV di-photon mass range) under 3\%, the significance could exceed the 5$\sigma$ threshold, indicating a strong potential for observation at the discovery level using Belle-II's existing dataset. 

Finally, also the possibility to produce para-TM using other processes, i.e. associated production ($e^+e^- \to TM \gamma \to 3 \gamma$), sensitive at the 2$\sigma$ level, and photon-photon fusion process with visible $e^+e^-$ final states, exceeding 5$\sigma$ expected significance, is briefly outlined.

\section{Acknolegments}
The authors are grateful to M. Raggi and to D. Redigolo for useful discussions and suggestions. SP, in particular, would like to thank H. Lamm for support and insightful discussions on True Muonium.

\bibliography{apssamp}

\end{document}